\documentclass[aps,prd,reprint,amsmath,amssymb,twocolumn,preprintnumbers,superscriptaddress,nofootinbib,letterpaper]{revtex4}

\usepackage{slashed}
\usepackage{bm}
\usepackage{latexsym,amssymb,amsmath,float,url,mathrsfs}
\usepackage{latexsym}
\usepackage{graphicx}
\usepackage{hyperref}
\usepackage[dvipsnames]{xcolor}
\usepackage{tabularx}

\hypersetup{
     colorlinks   = true,
     citecolor    = violet,
     urlcolor     = violet,
     linkcolor    = violet
}

\preprint{MIT-CTP/5020}

\begin{document}

\title{GeV-Scale Thermal WIMPs: Not Even Slightly Dead}

\author{Rebecca K.\ Leane}
\thanks{{\scriptsize Email}: \href{mailto:rleane@mit.edu}{rleane@mit.edu}; {\scriptsize ORCID}: \href{http://orcid.org/0000-0002-1287-8780}{0000-0002-1287-8780}}
\affiliation{Center for Theoretical Physics, Massachusetts Institute of Technology, Cambridge, MA 02139, USA}

\author{Tracy R. Slatyer}
\thanks{{\scriptsize Email}: \href{mailto:tslatyer@mit.edu}{tslatyer@mit.edu}; {\scriptsize ORCID}:
\href{http://orcid.org/0000-0001-9699-9047}{0000-0001-9699-9047}}
\affiliation{Center for Theoretical Physics, Massachusetts Institute of Technology, Cambridge, MA 02139, USA}

\author{John F. Beacom}
\thanks{{\scriptsize Email}: \href{mailto:beacom.7@osu.edu}{beacom.7@osu.edu}; {\scriptsize ORCID}:
\href{http://orcid.org/0000-0002-0005-2631}{0000-0002-0005-2631}}
\affiliation{Center for Cosmology and AstroParticle Physics (CCAPP), Ohio State University, Columbus, OH 43210, USA}
\affiliation{Department of Physics, Ohio State University, Columbus, OH 43210, USA}
\affiliation{Department of Astronomy, Ohio State University, Columbus, OH 43210, USA}

\author{Kenny C. Y. Ng}
\thanks{{\scriptsize Email}: \href{mailto:chun-yu.ng@weizmann.ac.il}{chun-yu.ng@weizmann.ac.il}; {\scriptsize ORCID}:
\href{http://orcid.org/0000-0001-8016-2170}{0000-0001-8016-2170}}
\affiliation{Department of Particle Physics and Astrophysics, Weizmann Institute of Science, Rehovot 76100, Israel}

\date{\today}

\begin{abstract}
Weakly Interacting Massive Particles (WIMPs) have long reigned as one of the leading classes of dark matter candidates. The observed dark matter abundance can be naturally obtained by freezeout of weak-scale dark matter annihilations in the early universe. This ``thermal WIMP'' scenario makes direct predictions for the total annihilation cross section that can be tested in present-day experiments.  While the dark matter mass constraint can be as high as $m_\chi\gtrsim100$~GeV for particular annihilation channels, the constraint on the \emph{total} cross section has not been determined. We construct the first model-independent limit on the WIMP total annihilation cross section, showing that allowed combinations of the annihilation-channel branching ratios considerably weaken the sensitivity. For thermal WIMPs with $s$-wave $2\rightarrow2$ annihilation to visible final states, we find the dark matter mass is only known to be $m_\chi\gtrsim20$~GeV. This is the strongest largely model-independent lower limit on the mass of thermal-relic WIMPs; together with the upper limit on the mass from the unitarity bound ($m_\chi\lesssim 100$ TeV), it defines what we call the ``WIMP window''. To probe the remaining mass range, we outline ways forward.
\end{abstract}

\maketitle

\section{Introduction}
\label{sec:intro}

A leading candidate for dark matter (DM) is a Weakly Interacting Massive Particle (WIMP) that is a thermal relic of the early universe~\cite{Bergstrom:2000pn,Bertone:2004pz}. For masses above $\sim 1$ keV, such a particle behaves as Cold Dark Matter (CDM)~\cite{Steigman:1984ac}, with dynamics governed by purely gravitational interactions. CDM is in excellent agreement with all large-scale observations of the universe, although there are some persistent discrepancies on smaller scales, where baryonic physics is also important.

The defining feature of the thermal WIMP is that its relic abundance is naturally explained by the freezeout process~\cite{Steigman:2012nb} with a weak-scale cross section, $\Omega_\chi h^2 \sim 0.1$ pb$~\times~c /\langle\sigma v \rangle$, where $\Omega_\chi h^2\approx 0.12$~\cite{Patrignani:2016xqp} is the DM density and $\langle\sigma v \rangle$ is the thermally averaged annihilation cross section. The weak cross section would explain why non-gravitational interactions of DM have so far evaded detection. In many Beyond the Standard Model (BSM) theories, there are WIMP candidates that naturally appear around the weak scale. While a simple thermal WIMP is by far not the only possibility for DM, it is a well-motivated scenario that must be decisively tested.

Although there are strong limits on WIMP scattering from direct detection and WIMP production from colliders~\cite{Carpenter:2012rg, Carpenter:2013xra, Petrov:2013nia, Bell:2012rg, Bell:2015rdw, Birkedal:2004xn, Gershtein:2008bf, Goodman:2010ku, Crivellin:2015wva, Petriello:2008pu, Berlin:2014cfa, Lin:2013sca, Fox:2011pm, Bell:2015sza, Bai:2015nfa, Autran:2015mfa,Gupta:2015lfa,Ghorbani:2016edw,Abercrombie:2015wmb,Escudero:2016gzx,Amole:2015pla, Amole:2016pye, Fu:2016ega, Akerib:2016lao, Aprile:2016swn,Aprile:2017iyp, Cui:2017nnn, PhysRevLett.118.021303, Aaboud:2016qgg, Aaboud:2016obm, Aaboud:2017dor, Aaboud:2017uak, Aaboud:2017yqz, Aaboud:2017bja, Aaboud:2017rzf, Aaboud:2017phn, Sirunyan:2017onm, Sirunyan:2017hci, Sirunyan:2017hnk, Sirunyan:2017xgm, Sirunyan:2018gka, Sirunyan:2018wcm,Baer:2014eja,Roszkowski:2017nbc}, these have not been shown to deliver model-independent sensitivity to generic thermal WIMP scenarios.  The branching ratios, coupling types and signals are model-dependent, and so the lack of observations may just be due to such features. For example, there can be interference effects, momentum suppression, or velocity suppression, that make the direct detection and collider cross sections small even when the total annihilation cross section is not.  In other words, there is no well-specified target scale for these experiments. There is for annihilation.  Thus we focus on annihilation, and especially on the total cross section.

While the thermal WIMP hypothesis specifies the total cross section, {\it there are no model-independent predictions for the annihilation branching ratios to specific final states.}  Thus, although there are many constraints on individual annihilation channels from various indirect detection searches, some of which even reach below the thermal WIMP cross section, a consistent and conservative interpretation of the data in the context of the generic thermal WIMP is surprisingly lacking; we aim to address this in this work. 

To decisively test thermal WIMPs, the sensitivity on the total cross section considering all possible SM final states needs to be calculated.  It must be tested for mass scales from $\sim$keV (minimum for CDM) to $\sim$100 TeV (unitarity)~\cite{Griest:1989wd,Blum:2014dca}. We ignore invisible final states as by definition, they cannot be tested with indirect detection.  We also ignore neutrino final states, which are categorically more difficult to probe~(though not impossible~\cite{Beacom:2006tt,Yuksel:2007ac,Aartsen:2017ulx,ElAisati:2017ppn}.)  The most robust limits on thermal WIMP annihilation come from three sources.  Below about 10 GeV, the strongest are from \textit{Planck} measurements of the CMB~\cite{Ade:2015xua}, which is sensitive to the total ionizing energy injected.  Electron and photon primary final states inject almost all their energy into electromagnetically (EM) interacting particles, while others inject at least about 25$\%$. Due to the precision of \textit{Planck}, these limits have low relative theoretical uncertainties, and are very robust. Furthermore, they do not depend on present-day annihilation rates. \textit{Planck} limits apply to much lower DM masses with linear improvement down to the $\mathcal{O}$(keV) scale. In addition, BBN does not allow a generic WIMP below about 10 MeV~\cite{Boehm:2013jpa,Nollett:2013pwa,Nollett:2014lwa}. (For DM masses below the electron mass, there are also strong limits on gamma rays from annihilation in the Milky Way and beyond~\cite{Mack:2008wu,Essig:2009jx}.)  Above about 10 GeV, \textit{Fermi} measurements of Dwarf Spheroidal Galaxies of the Milky Way, and the Alpha Magnetic Spectrometer (AMS) measurements of cosmic-rays are the strongest robust limits. \textit{Fermi} has the best sensitivity to photon-rich final states but not leptons~\cite{Ackermann:2015zua, Fermi-LAT:2016uux}, while AMS is most sensitive to leptons but not photons~\cite{Aguilar:2014mma, Accardo:2014lma}. Particularly for AMS, there are considerable astrophysical uncertainties, but their effect can be mitigated by making conservative choices.

In this paper, we perform the first calculation of the limit on the total WIMP annihilation cross section, combining data on all kinematically allowed final states from {\it Fermi} gamma-ray observations of Dwarf Spheroidal Galaxies, AMS-02 positron flux measurements, and {\it Planck} CMB energy measurements. Together with the unitarity bound, this defines the ``WIMP window,'' as discussed further below. In Sec.~\ref{sec:excluded}, we discuss when the generic thermal WIMP can be considered excluded. In Sec.~\ref{sec:method}, we describe our general approach to set a lower limit on the thermal WIMP mass. We then provide specific details for setting limits with \textit{Planck}, \textit{Fermi} and AMS in Sec.~\ref{sec:planck},~\ref{sec:fermi},~\ref{sec:ams} respectively. We discuss our results in Sec.~\ref{sec:results}, and important progress for testing the WIMP paradigm in Sec.~\ref{sec:outlook}, before concluding in Sec.~\ref{sec:conclusion}.

\section{When is a thermal WIMP excluded?}
\label{sec:excluded}

Searches for dark matter annihilation products have set strong limits in certain cases, requiring that the dark matter mass be $m_{\chi} \gtrsim 100$ GeV if annihilation proceeds solely to $b$ quarks (\textit{Fermi}), $\tau$ leptons (\textit{Fermi}), or electrons (AMS). These are only upper limits, and only apply when the limit is dominated by favorable final states. How do we quantify, more generally, when the minimal thermal WIMP is excluded?

To meaningfully combine limits on all final states, we use the simple point that branching fractions of DM must add to 100$\%$. 
At a particular mass, thermal-relic WIMP is excluded if, for the standard total cross section, no combination of final-state branching ratios is in accord with all constraints simultaneously. More generally, we define the limit on the total cross section as the largest value for which all constraints are satisfied.

A limit on favorable final states corresponding to $m_\chi\gtrsim$~100 GeV does not mean that lower masses are uninteresting.  In fact, it's only for $m_\chi \lesssim 100$ GeV that we then start to have the sensitivity to actually test the general thermal-WIMP hypothesis.  This is when making informative statements (by combining branching ratios) starts, not ends. For higher masses, we learn nothing about a generic WIMP, only that better sensitivity is needed.

When quantifying when the WIMP hypothesis is excluded using experimental data, we must consider the following points.

\subsection{Conservative Values for Experimental Inputs}
First, in order to make a decisive statement that certain masses are excluded, we must choose conservative values for the various parameters that influence the results. 
While there are large uncertainties present in astrophysical searches, after conservative parameter choices, the residual uncertainties are low. This is discussed in Sec.~\ref{sec:uncertainties}.

\subsection{Precision of the Thermal Target Value}
Second, we must ask how precisely the thermal relic cross section target is known ~\cite{Steigman:2012nb}. The thermally averaged relic cross section can be expanded in partial waves,
\begin{equation}
 \langle\sigma v\rangle = \langle a+bv^2+cv^4+... \rangle = a+ \frac{3}{2}\frac{b}{x}+\frac{15}{8}\frac{c}{x^2}+...
\end{equation}
where $x=m/T$ ($x\sim20$ at freezeout), $v$ is the relative DM velocity, $a$ is the velocity independent $s$-wave contribution, $b$ contains the leading $p$-wave contribution, and $c$ contains the leading $d$-wave contribution. 

In the standard calculation, the uncertainty in the thermal relic cross section is very small (at the percent level)~\cite{Steigman:2012nb}, arising from the uncertainty on the measurement of the matter density $\Omega_m h^2$~\cite{Patrignani:2016xqp}. However, including order $v^2$ contributions to $\langle\sigma v\rangle$ during freezeout can provide a $5-10\%$ correction to the pure $s$-wave piece, based on the estimate that freezeout happens around $x\sim20$, i.e., $v^2 \sim 0.1$. This uncertainty goes in only one direction, in that increasing the $\mathcal{O}$($v^2$) piece of $\langle\sigma v\rangle$ means a smaller $\mathcal{O}$($v^0$) cross section is required to get the correct relic density. The exact relation is model-dependent.

The set of reasonable cases considered must include the vanilla thermal WIMP ($s$-wave $2\rightarrow2$ annihilation into visible final states) with a standard cosmological history.  Thus, absent new data, the general lower limit cannot be stronger than what we calculate here.  It can be weaker, which strengthens our point that GeV-scale thermal WIMPs have not yet been adequately probed.

For example, in the case of Majorana DM, the DM annihilation is helicity suppressed, scaling like $(m_{\rm SM}/m_\chi)^2$, where $m_{\rm SM}$ is the mass of the DM annihilation product. For light final states, this renders the $s$-wave annihilation sub-dominant, and so annihilation may proceed dominantly in the $p$-wave. This leads to a difference in the late time annihilation cross section and the cross section at freezeout. More generally, if there are other annihilation channels that are relevant at freezeout but not for indirect detection (or vice versa), the cross section target can be affected~\cite{Griest:1990kh}. For example, if there are multiple DM particles, there can be co-annihilation with an unstable species that has decayed by the time indirect probes become relevant.
Considering the most generic scenario leads to a certain limit; considering a set of more general scenarios --- which must include the generic one --- can only lead to a weaker limit.

\subsection{WIMP Signal Generation Uncertainties}

Third, it is important to consider uncertainties in generating a WIMP signal. We use {\sc Pythia8.2}~\cite{Sjostrand:2014zea} to generate the energy spectra of stable secondary particles produced by DM annihilation, for the various SM final states. For center of mass energies $\sim 1-10$ GeV (DM masses $\sim0.5-5$~GeV), there are hadronic resonances that may render the \textsc{Pythia} calculations less reliable~\cite{Sjostrand:2014zea, Ezhela:2003pp}. While we will show \textsc{Pythia}-based results in this mass range, we will also present a general argument, independent of the details of the \textsc{Pythia} output, that \textit{Planck} limits exclude the thermal relic cross section over this range (see Sec.~\ref{sec:energyinj}). When the final states are dominated by muons, electrons, and photons, it is possible to use analytic expressions for the energy spectra, which we use below 1 GeV. 

There are also uncertainties in the modeling of stable secondary particles from the various annihilation channels (reflected in discrepancies between \textsc{Pythia} and other event generators~\cite{Cirelli:2010xx, Bahr:2008pv}). However, the scenarios with largest uncertainty do not contribute substantially to our least-constrained combination of annihilation channels, and so we expect the effect of these uncertainties to be small. We also note that the additional radiative muon decays pointed out in Ref.~\cite{Scaffidi:2016ind} do not affect our results within our precision goal. 

\subsection{Choice of Statistical Significance}

Fourth, the choice of statistical significance changes the upper limit on the annihilation cross section. In astrophysical searches, it is common to present limits at 95$\%$ C.L. Similarly, we present our results to 95$\%$ C.L., but show in Sec.~\ref{sec:uncertainties} that increasing to $(1-10^{-7})\times100\%$ C.L. ($5\sigma$) weakens limits by a factor of between about $\sim$1.5--10, depending on the DM mass. Overall, for a fixed C.L., we keep our calculations within the precision goal of 50$\%$. That is, we neglect some uncertainties that we expect to affect the limit by less than 50$\%$.

\subsection{Degree of Belief}
\label{sec:belief}
Even in light of the care we have taken to be conservative in all choices, one might not accept that a WIMP is ruled out if its maximum allowed cross section is ``just below" the thermal-relic prediction.  Accordingly, we later note the mass limits that result if one requires the clearance to be a factor of 2, which is arguably reasonable, or a factor of 10, which is clearly excessive.  These lead to somewhat smaller lower limits on the WIMP mass, strengthening our point that the often-quoted $\sim 100$ GeV for particular final states is too optimistic for a general limit.

\section{General Methodology}
\label{sec:method}

We calculate the largely model-independent upper limit on the thermal WIMP cross section, correctly combining inputs from leading astrophysical and cosmological experiments. For \textit{Fermi} and AMS, this limit is not just a linear scaling of individual channel limits from each experiment. This is because the limits are set on the gamma-ray or positron spectral energy distribution (SED) respectively. Introducing mixed final states will change the signal spectra for a given cross section, modifying the bin-by-bin energy flux, which is what determines the limit. For the CMB, while existing limits above $\sim$5 GeV scale linearly, in order to extend the limits to the sub-5-GeV DM mass range for general final states, we need to work in a regime where hadronic resonances are potentially important. We will present two methods for setting limits in this mass range.

All branching channels are considered where kinematically allowed (except neutrinos). This includes annihilation to electrons $e$, muons $\mu$, taus $\tau$, $b$-quarks $b$, gamma rays $\gamma$, gluons $g$, $W$-bosons $W$,  $Z$-bosons $Z$, Higgs bosons $H$, and light quarks that are grouped into the channel $q=u,d,s,c$. We scan over all branching fractions, with 5$\%$ incremental difference for each channel, and 1 GeV incremental difference for DM mass. 

For all combinations of branching ratios, gamma-ray and positron energy spectra are generated using {\sc Pythia8.2}~\cite{Sjostrand:2014zea}. To generate spectra below 10 GeV center-of-mass energies, we run {\sc Pythia} with back-to-back beams with \mbox{\tt Beams:frameType=2}, rather than the standard center-of-mass beam mode. We also increase the available phase space with \mbox{\tt PhaseSpace:mHatMin = 1.0}. We use these spectra to derive limits using \textit{Fermi}, AMS and \textit{Planck} data, as described in the following sections. Once the limit on the annihilation cross section is below the thermal relic value~\cite{Steigman:2012nb}, the particular DM mass is excluded. Note that while single annihilation channels, or limited combinations, have been previously fit to the multiple experiments simultaneously~(e.g.,~\cite{Cirelli:2008pk,Lopez:2015uma,Cuoco:2016eej,Cuoco:2017rxb,Eiteneuer:2017hoh,Jia:2017kjw,Arcadi:2017vis}), a limit on the total annihilation cross section has not previously been constructed.

\section{Planck CMB Limits}
\label{sec:planck}

Anisotropies of the CMB provide powerful insight into physical processes present during the cosmic dark ages. Any injection of ionizing particles, including those from DM annihilation, modifies the ionization history of hydrogen and helium gas, perturbing CMB anisotropies. Measurements of these anisotropies therefore provide robust constraints on production of ionizing particles from DM annihilation products. The most sensitive measurements to date are by \textit{Planck}~\cite{Ade:2015xua}, superseding earlier measurements by WMAP~\cite{2013ApJS..208...20B}.

\begin{figure}[t]
\centering
\includegraphics[width=1.02\columnwidth]{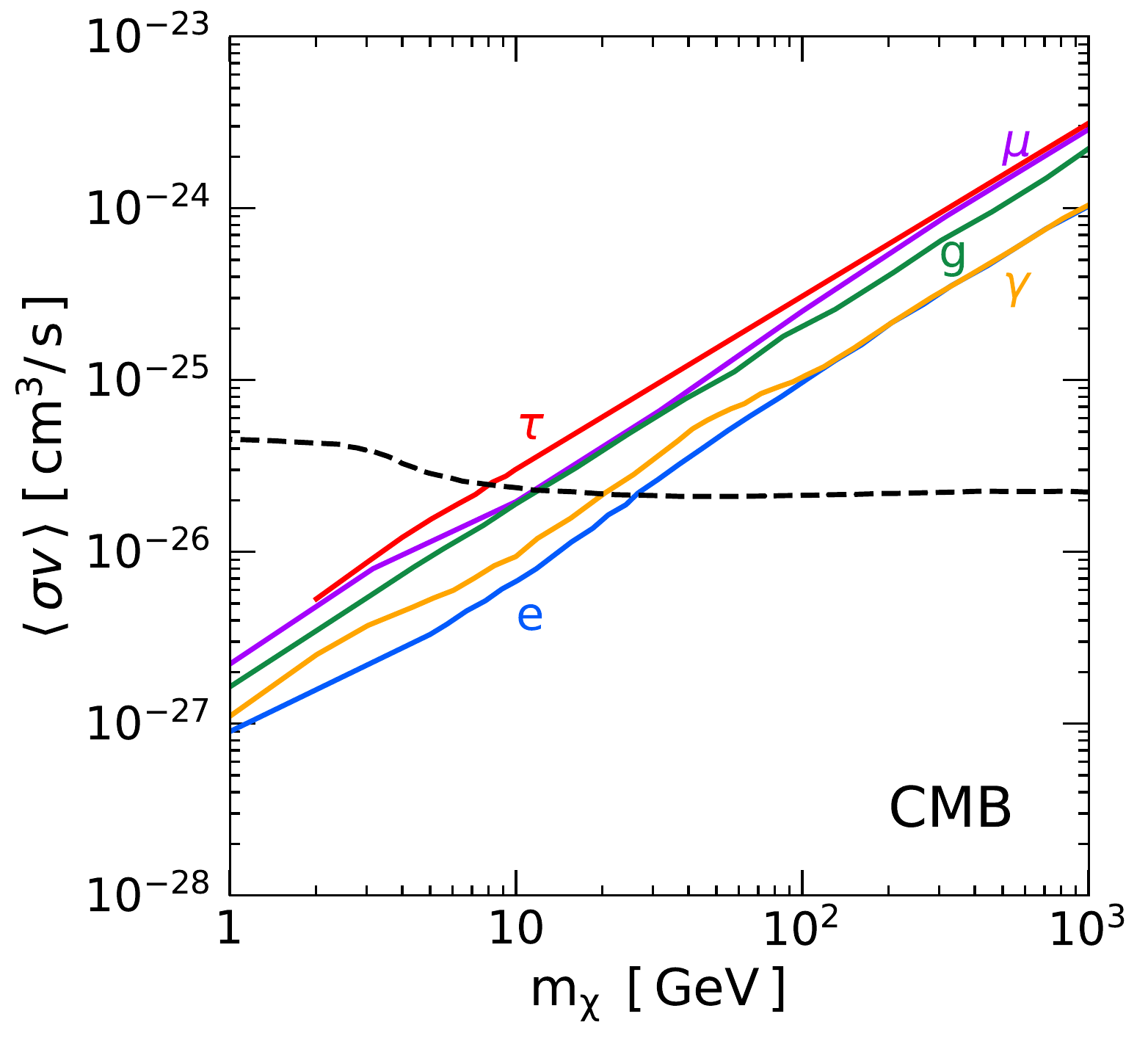}
\caption{\textit{Planck} CMB limits at 95$\%$ C.L. for DM annihilation 100$\%$ to individual channels: electrons (blue), muons (purple), taus (red), gluons (green), gamma rays (orange). Light quarks and $b$-quarks overlap with the gluon line, so are not shown for clarity. Thermal relic cross section is the black dashed line~\cite{Steigman:2012nb}.}  
\label{fig:CMBlimit}
\end{figure}

\begin{figure}[t]
\centering
\includegraphics[width=0.97\columnwidth]{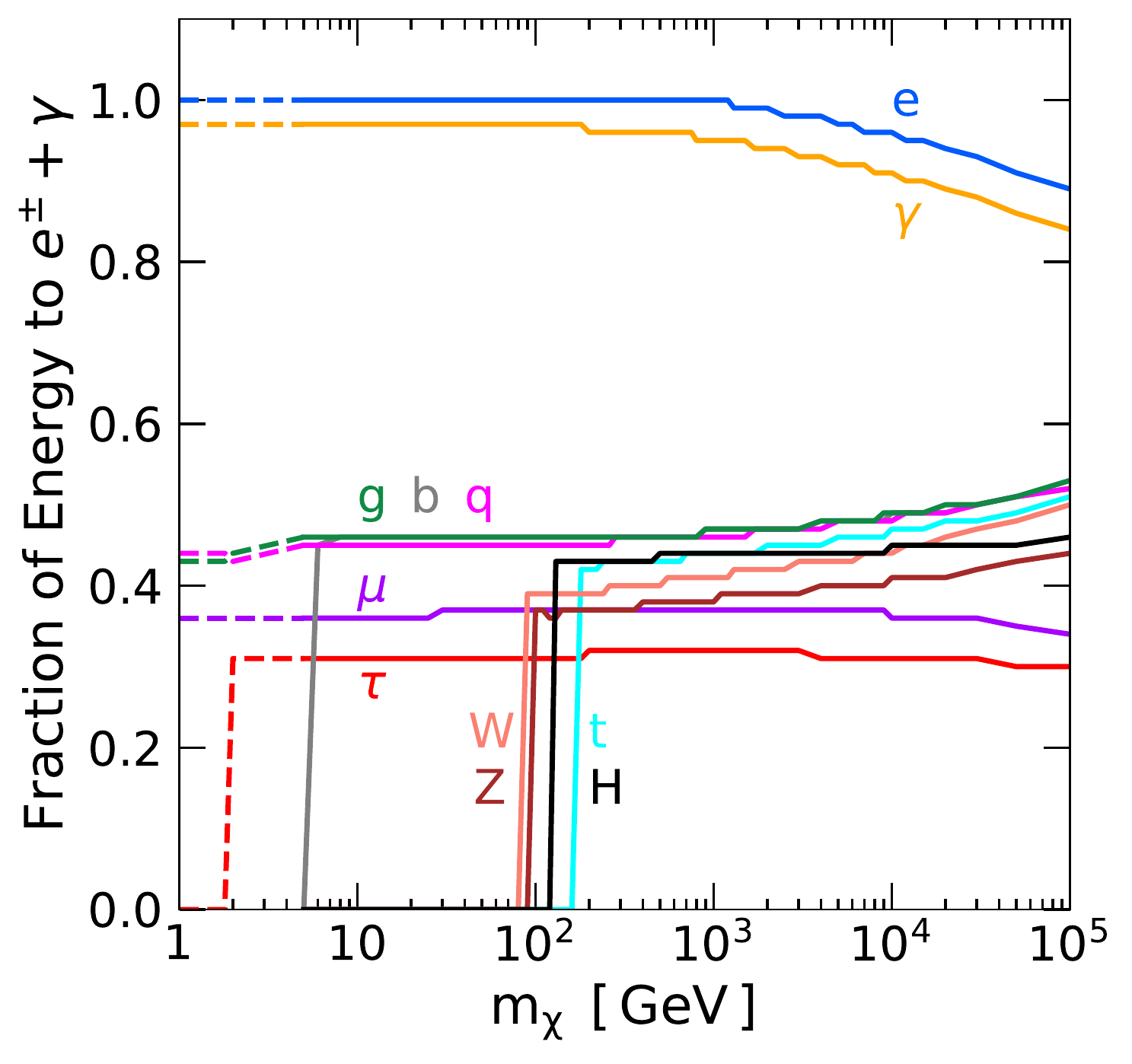}
\caption{Fraction of energy from primary DM annihilation states into EM interacting products (electrons + positrons + photons). Shown are electrons $e$, muons $\mu$, taus $\tau$, light quarks $q$, $b$-quarks $b$, gluons $g$, $W$-bosons $W$, $Z$-bosons $Z$, Higgs bosons $H$, and top-quarks $t$. The dashed line is the hadronic resonance region.}  
\label{fig:CMBenergyfrac}
\end{figure}

\subsection{Energy Injection from Annihilating DM}
\label{sec:energyinj}
The power deposited by DM annihilation, controlled by the parameter
\begin{equation}
 p_{\rm ann}=f_{\rm eff}\frac{\langle\sigma v\rangle}{m_\chi},
\end{equation}
determines the strength of the CMB limit. Here $\langle\sigma v\rangle$ is the thermally averaged DM annihilation cross section and $m_\chi$ is the DM mass. We calculate the weighted efficiency factor $f_{\rm eff}$ by integrating our electron/positron and photon energy spectra from {\sc Pythia} over the $f_{\rm eff}^{e^\pm,\gamma}(E)$ curves calculated in Ref.~\cite{Slatyer:2015jla},
\begin{equation}
 f_{\rm eff}(m_\chi)=\frac{1}{2m_\chi}\int_0^{m_\chi}\left(f_{\rm eff}^{e^\pm}\frac{dN}{dE_{e^\pm}}+f_{\rm eff}^\gamma\frac{dN}{dE_\gamma}\right)EdE.
\end{equation}
Following Ref.~\cite{Slatyer:2015jla}, we neglect the contribution to energy deposition from protons and antiprotons; generally only a small fraction of the total energy of the annihilation products goes into $p\bar{p}$ production, and protons and antiprotons also deposit energy less efficiently than electrons, positrons, and photons~\cite{Weniger:2013hja}. Including these contributions would slightly strengthen the constraints.
From \textit{Planck} data, the 95$\%$ C.L. limit on $p_{\rm ann}$ is~\cite{Ade:2015xua}
\begin{equation}
 f_{\rm eff}\frac{\langle\sigma v\rangle}{m_\chi} < 4.1\times10^{-28}\, \rm cm^3/s/GeV.
 \label{eq:fefflim}
\end{equation}
Figure~\ref{fig:CMBlimit} shows the single-channel limits on the cross section from the CMB. Below 5 GeV DM mass, as there is extra uncertainty in the {\sc Pythia} spectra, we also present arguments for the thermal WIMP exclusion based on generic arguments about the efficiency and energy injection rate, as discussed below.

\subsection{Energy Injection Fractions}

Figure~\ref{fig:CMBenergyfrac} shows the fraction of power proceeding into EM channels (electrons, positrons, and photons) is quite stable as a function of DM mass, and is 26$\%$ or higher for all DM masses between 5 GeV and 10 TeV and all (non-neutrino) final states. (Note these are the final branching ratios after all the unstable particles decay, not the direct branching ratios into electrons, positrons, and photons.)

For annihilation to electrons and muons, this statement is fairly trivial, as the fraction is 100\% for electrons or simply determined by 3-body kinematics for muons (resulting in roughly 1/3 of the energy going into electrons and the other 2/3 into neutrinos); the only subtlety is at high masses, where electroweak corrections allow neutrinos to be produced even for the $e^+ e^-$ final state. For other final states with hadronic decays, the decays will typically proceed through either neutral or charged pions. Neutral pions decay to photons with a nearly 100\% branching ratio, while charged pions decay first to a muon and antineutrino (or antimuon and neutrino), with the muon then decaying to an electron + neutrino + antineutrino. In the latter process, the muon receives $\sim 80\%$ of the pion rest energy, and then the electron carries away roughly 1/3 of the muon energy, so $\sim 25\%$ of the pion's energy is carried by the final-state electron.

From these simple arguments, we would expect the fraction of power into EM channels to vary between $\sim 25\%$ and 100$\%$ for non-neutrino final states, in agreement with Fig.~\ref{fig:CMBenergyfrac}. There is no reason to expect this argument to break down for DM masses below 5 GeV, although the branching ratio into hadronic versus leptonic final states may change rapidly when the DM mass becomes close to a hadronic resonance. For hadronic final states, we furthermore expect that the energy of the produced photons/electrons will peak no lower than a $\mathcal{O}(1)$ fraction of the pion mass (and the QCD scale); likewise, muon decays will typically produce electrons with $\mathcal{O}(10$--100) MeV energies.

We therefore robustly expect that for DM masses between $\sim 100$ MeV and 5 GeV, at least 25\% of the DM rest energy should go into producing photons, electrons and positrons with energies above 5 MeV. Comparing with the dashed lines in Fig.~\ref{fig:CMBenergyfrac}, which are results from \textsc{Pythia}, confirms this argument in the hadronic resonance region. However as \textsc{Pythia} carries extra uncertainty in this regime, we use this estimate as a conservative cross check to set a robust constraint on light DM annihilation.

For electron/positron/photon energies above 5 MeV, the minimum value of $f_\text{eff}^{e^\pm,\gamma}$ is $0.32$. Thus we expect $f_\text{eff}$ for any 2-body SM final state other than neutrinos to exceed $f_\text{min} = 0.25\times 0.32 \approx 0.08$ for DM masses in the 100 MeV -- 5 GeV window (for masses below this window, the results for direct annihilation to electrons/positrons/photons can be used). (As a cross-check that this is conservative, the minimum $f_\text{eff}$ value for DM masses above 5 GeV is 0.12 for the same set of channels; realistically all the electrons/positrons/photons will not be concentrated at the energies that minimize $f_\text{eff}$.)

Taking the \textit{Planck} $95\%$ confidence limit in Eq.~(\ref{eq:fefflim}), this conservative $f_\text{min}$ model-independently implies 
\begin{equation}
\langle \sigma v \rangle <  5.1\times 10^{-27}\,\times\left(\frac{m_\chi}{\rm GeV}\right) \,{\rm cm^3/s}
\label{eq:cmbhadronic}
\end{equation}
for DM masses below 5 GeV, which definitively excludes the $s$-wave thermal relic cross section in this mass range.

\section{Fermi-LAT Dwarf Spheroidal Gamma-Ray Limits}
\label{sec:fermi}

Dwarf Spheroidal Galaxies (dSphs) of the Milky Way are one of the best DM signal targets, as according to kinematic data they are DM dense with low background. \textit{Fermi} has set limits on gamma-ray fluxes from discovered dSphs~\cite{Ackermann:2015zua, Fermi-LAT:2016uux}, with no conclusive DM signal. DSphs provide some of the strongest limits on DM annihilating to any photon-rich final states, such as gamma-ray lines or hadronic final states.

\subsection{Fit to data}

To set limits on photons from mixed final states, we follow the official \textit{Fermi} analysis on \texttt{Pass 8} LAT data~\cite{Fermi-LAT:2016uux} and consider a total of 41 dwarf galaxies, both kinematically confirmed and likely galaxies\footnote{The bin-by-bin likelihoods for each dwarf can be downloaded from  \href{http://www-glast.stanford.edu/pub_data/1203/}{http://www-glast.stanford.edu/pub$\_$data/1203/}}.
Where provided, we use the measured $J$-factor and uncertainty. This is for 19 dwarf galaxies: Bootes I, Canes Venatici I, 
Canes Venatici II, Carina, Coma Berenices, Draco, Fornax, Hercules, Leo I, Leo II, Leo IV, Leo V, Reticulum II, Sculptor, Segue 1, 
Sextans, Ursa Major I, Ursa Major II, and Ursa Minor.
For the remaining 22 galaxies not named above, spectroscopic $J$-factors are unavailable, and following Ref.~\cite{Fermi-LAT:2016uux} we use the predicted $J$-factors with a nominal uncertainty of 0.6 dex. 
Dwarfs we consider in this category are Bootes II, Bootes III, Draco II, Horologium I, Hydra II, Pisces II, 
Triangulum II, Tucana II, Willman 1, Columba I, Eridanus II, Grus I, Grus II, Horologium II, Indus II, Pegasus III, 
Phoenix II, Pictor I, Reticulum III, Sagittarius II, Tucana III, and Tucana IV.

Note that four of these galaxies (Reticulum II, Tucana III, Tucana IV, Indus II) have shown a local $\sim2\sigma$ excess in gamma rays~\cite{Fermi-LAT:2016uux}, which may be attributed to DM. However, this is not globally significant, so we do not fit these excesses to a DM signal, and instead treat the measurements as exclusion bounds.

For each of these dwarf galaxies, \textit{Fermi} provides the likelihood curves as a function of the integrated energy flux,
\begin{equation}
\Phi_E = \frac{\langle \sigma v \rangle}{8\pi m_{\chi}^2} \left[ \int_{E_{\rm min}}^{E_{\rm max}} E \frac{dN}{dE} dE \right] J_i\,,
\label{eq:Eflux}
\end{equation}
where $J_i$ is the $J$-factor for dwarf $i$. Following Ref.~\cite{Ackermann:2015zua}, we treat the energy bins as independent, and obtain the full likelihood $\mathcal{L}_i\left({\mu} \vert \mathcal{D}_i\right)$, which is a function of the model parameters $\mu$ and data $D_i$, by multiplying the likelihoods for each for the 41 dwarfs together. The uncertainty in the $J$-factor is included as a nuisance parameter on the global likelihood, modifying the likelihood,
\begin{align}
&\tilde{\mathcal{L}}_i\left({\mu}, J_i \vert \mathcal{D}_i\right) = \mathcal{L}_i\left({\mu} \vert \mathcal{D}_i\right) \\&\times \frac{1}{\ln(10) J_i \sqrt{2\pi} \sigma_i} e^{-\left( \log_{10}(J_i) - \overline{\log_{10}(J_i)} \right)^2/2 \sigma_i^2} \nonumber
\label{eq:like}
\end{align}
as per the profile likelihood method \cite{Rolke:2004mj}. For $\overline{\log_{10}(J_i)}$ and $\sigma_i$, we use the values provided in \cite{Fermi-LAT:2016uux} for a Navarro-Frenk-White profile \cite{Navarro:1996gj}. The likelihood is maximized to produce an upper limit on the annihilation cross section at $95\%$ C.L. 

\begin{figure}[t]
\centering
\includegraphics[width=\columnwidth]{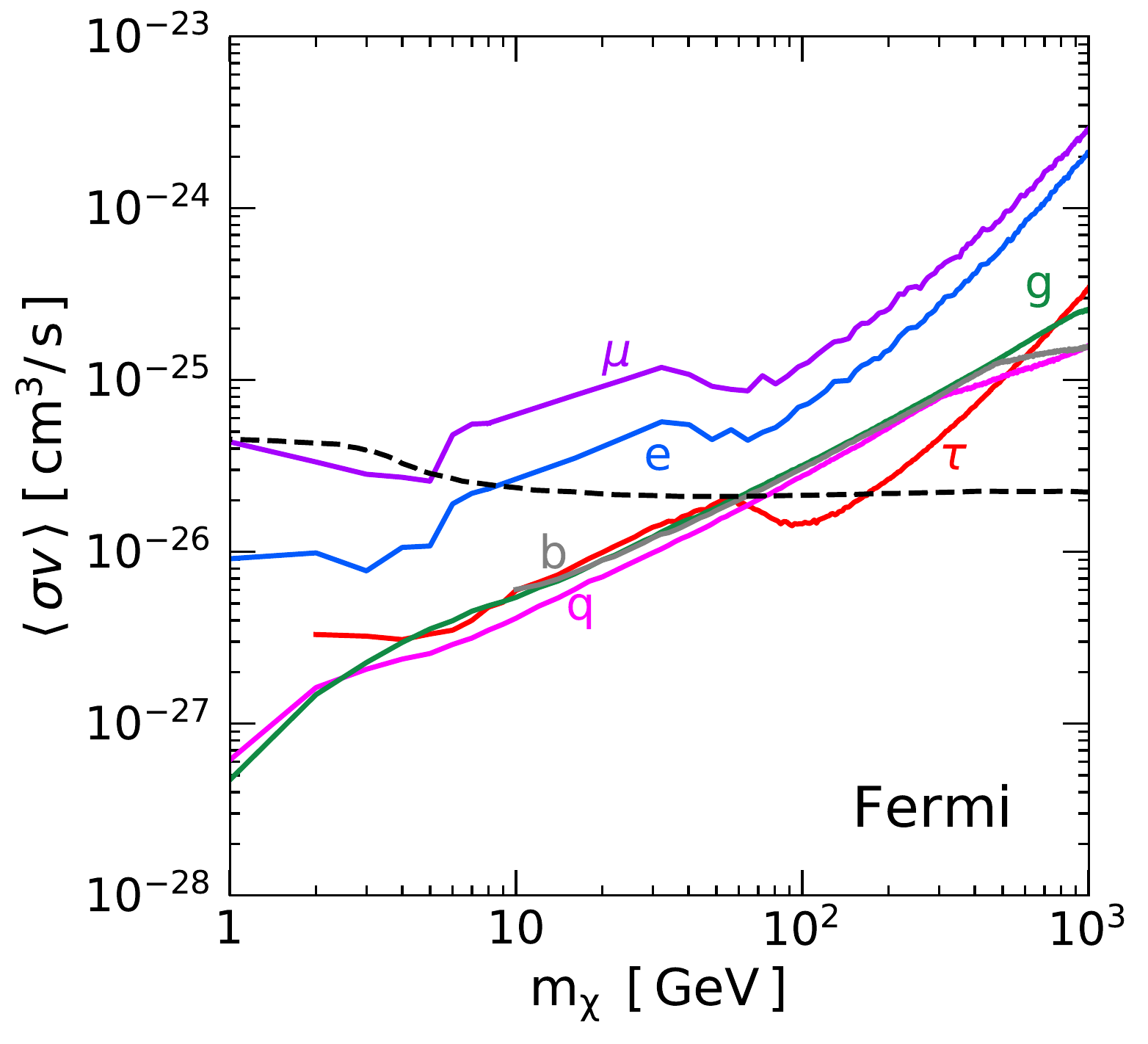}
\caption{\textit{Fermi}-LAT limits at 95$\%$ C.L. for DM annihilation 100$\%$ to individual channels: electrons (blue), muons (purple), taus (red), $b$-quarks (gray), gluons (green), and light quarks $q=u,d,s,c$ (magenta). Thermal relic cross section is the black dashed line~\cite{Steigman:2012nb}.}  
\label{fig:Fermilimit}
\end{figure}

Figure~\ref{fig:Fermilimit} shows our limits for the 100$\%$ branching fraction scenario. For the channels for which results are shown in Ref.~\cite{Fermi-LAT:2016uux}, our results are comparable. \textit{Fermi} gamma-ray line searches prohibit a large branching into gamma rays~\cite{PhysRevD.88.082002,Ackermann:2015lka}. Note that for TeV DM masses, limits from Imaging Atmospheric Cherenkov Telescopes (IACT) such as the High Energy Stereoscopic System (H.E.S.S.) are stronger~\cite{Abramowski:2011hc,Abramowski:2014tra,Abdalla:2016olq,Abdallah:2018qtu}, but do not probe the thermal relic cross section. H.E.S.S. gamma-ray line searches take over where \textit{Fermi} loses gamma-ray line sensitivity, and prohibits a large branching into gamma rays.

\begin{figure}[t]
\centering
\includegraphics[width=\columnwidth]{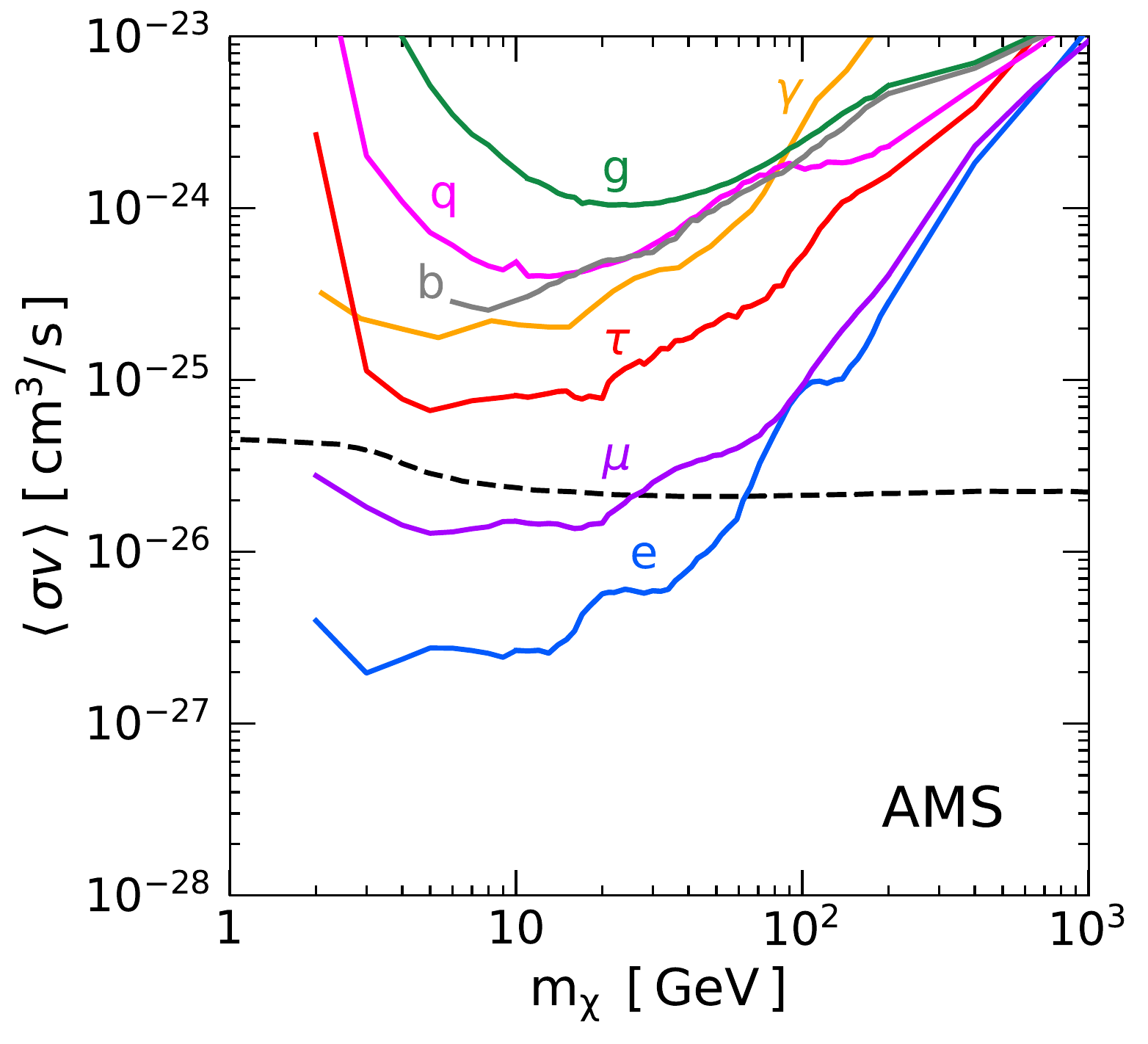}
\caption{Conservative AMS limits at 95$\%$ C.L. for DM annihilation 100$\%$ to individual channels: electrons (blue), muons (purple), taus (red), $b$-quarks (gray), gamma rays (orange), light quarks $q=u,d,s,c$ (magenta), and gluons (green). Thermal relic cross section is the black dashed line~\cite{Steigman:2012nb}.}
\label{fig:AMSlimit}
\end{figure}

\section{AMS-02 Positron Flux Limits}
\label{sec:ams}

 AMS measurements of electron and positron fluxes and fractions \cite{Aguilar:2014mma, Accardo:2014lma} provide the strongest constraints for electron and muon final states.
 We use the positron flux data to set a limit on DM annihilation to all final states. As we aim to set a robust exclusion on the WIMP annihilation cross section, and cosmic-ray propagation is not precisely understood, we must take conservative parameter values at every step.

\subsection{Cosmic-Ray Propagation}

To propagate our positron spectra generated with {\sc Pythia}, we use the cosmic-ray propagation program {\sc Dragon}~\cite{Evoli:2008dv, Evoli:2016xgn}.

The evolution of the number density $N_i$  of injected electrons and positrons is given by the diffusion equation,
\begin{eqnarray}
   \frac{ \partial N_i} { \partial t} &=&
      \vec{\nabla} \cdot \left( D \vec{\nabla} \right) N_i
      + \frac{\partial} { \partial p}\left(\dot{p}\right)N_i+ Q_i ( p, r, z)
   \nonumber \\
     &+&  \sum_{j>i} \beta n_\mathrm{gas} (r,z) \sigma_{ji} N_j
      - \beta n_\mathrm{gas} \sigma_i^{in}( E_k) N_i ~,
   \label{eqn:diff}
\end{eqnarray}
where the convection and diffusion in momentum space have been set to zero, as it does not largely affect the spectrum in the energy region of interest~\cite{Strong:2007nh}. In Eq.~(\ref{eqn:diff}), $D$ is the spatial diffusion coefficient, parametrized as
\begin{equation}
   D ( \rho, r, z) = D_0 {\rm e}^{|z|/z_t} \left( \frac{ \rho} 
      {\rho_0} \right)^\delta~,
\label{eqn:D}
\end{equation}
where $\rho = p / (Ze)$ is the rigidity of the charged
particle with $Z= 1$ for electrons and positrons. The diffusion is normalized by
$D_0$ at the rigidity $\rho_0 = 4$ GV. We assume the diffusion zone is axisymmetric with thickness $2 z_t$. In Eq.~(\ref{eqn:diff}), $\dot{p}$ accounts for the energy loss; 
$Q_i$ is the source term, where the DM source contribution is
\begin{equation}
   Q_\chi (p,r,z) = \frac{\rho_\chi^2(r)\langle\sigma v\rangle}{2m_\chi^2}\sum_f Br_f \frac{dN^f}{dE}.
\label{eqn:D}
\end{equation}
The effect of nuclei scattering with the gas is modeled by the second line in Eq.~(\ref{eqn:diff}).

The injected positrons are propagated using the model of Ref.~\cite{Evoli:2011id}. This sets $z_t=4$ kpc, $D_0=2.7\times10^{28}$ cm$^2$/s, $\delta=0.6$, but we take the local DM density to be the conservative $\rho=0.25$ GeV/cm$^3$, with an NFW profile.   We set the magnetic field at the Sun to be $B_\odot=8.9 \, \mu$G, which means that the local radiation field and magnetic field energy density is 3.1 eV/cm$^3$. As this is even higher than the conservative value of 2.6 eV/cm$^3$~\cite{Bergstrom:2013jra}, it leads to a higher energy loss rate for the positrons, which is the second-leading effect for CR propagation, behind the leading effect of the local DM density. As such, different choices of the other propagation parameters do not appreciably change the results. 

The most substantial energy-loss for charged cosmic rays below about 10 GeV is due to solar modulation. The largest measured value of the solar modulation potential during AMS's data taking period of $\Phi=0.6$ GV is taken~\cite{Cholis:2015gna}, and we employ the force-field approximation, which is valid for positron fluxes~\cite{Cholis:2015gna, Cavasonza:2016qem}.

\subsection{Fit to Data}

We assume the data measured by AMS do not have any DM source contributions. I.e., we do not assume the additional smooth ingredient over the astrophysical background measured by AMS is from DM, as the source of the additional ingredient is unknown. As such, rather than model backgrounds, we parameterize the total AMS measurements with a degree 6 polynomial function of variable $log(\rm energy)$ fit to $log(\rm flux)$. To set the limit we perform a likelihood ratio test, where the likelihood function is
\begin{equation}
\mathcal{L}(\theta)=\rm exp(-\chi^2(\theta)/2),
 \label{eq:AMSchi}
\end{equation}
where $\theta=\{\theta^1,\theta^2,...,\theta^n\}$ are parameters in the best fit polynomial function, and the $\chi^2(\theta)$ is given by
\begin{equation}
\chi^2(\theta)=\sum_i\frac{\left(f_i ^{th}(\theta)-f_i^ {data} \right)^2}{\sigma_i^2},
 \label{eq:AMSchi}
\end{equation}
where $f_i^{\rm th}$ is the prediction from the modeled background (the polynomial function), $f_i^{\rm data}$ is the central flux energy bin of the AMS data, and $\sigma_i$ is the uncertainty on the particular flux value. We sum over all the AMS energy bins measured, and add the systematic and statistical uncertainty in quadrature. We then allow the parameters of the function to float within 30$\%$ of their best fit values without DM (increasing the allowed values does not weaken constraints), that allows the function to absorb a DM signal if it is preferred over the additional smooth component. We increase the DM signal normalization until the functional fit of the background plus signal to the data produces a $\chi^2$ which has increased by 2.71, i.e.,
\begin{equation}
\chi_{\rm DM}^2=\chi^2+2.71.
 \label{eq:AMSchi2}
\end{equation}
This produces an 95$\%$ C.L. upper limit on the DM annihilation channel. 

Figure~\ref{fig:AMSlimit} shows our limits for the case of 100$\%$ branching fractions. We check that we can reproduce comparable results to similarly conservative scenarios from Ref.~\cite{Elor:2015bho, Cavasonza:2016qem}. Our results are more conservative than the weakest region of the bounds presented in Ref.~\cite{Bergstrom:2013jra}, which arises from taking the choices for propagation parameters that lead to the weakest limit (see Sec.~\ref{sec:uncertainties} for more details).

\section{Results and Discussion}
\label{sec:results}

We now present our results combining all limits from \textit{Planck} (Sec.~\ref{sec:planck}), \textit{Fermi} (Sec.~\ref{sec:fermi}) and AMS (Sec.~\ref{sec:ams}), using the method of combining all kinematically allowed branching fractions as described in Sec.~\ref{sec:method}.

\subsection{Limit on the Total Annihilation Cross Section}

\begin{figure}[t]
\centering
\includegraphics[width=1.05\columnwidth]{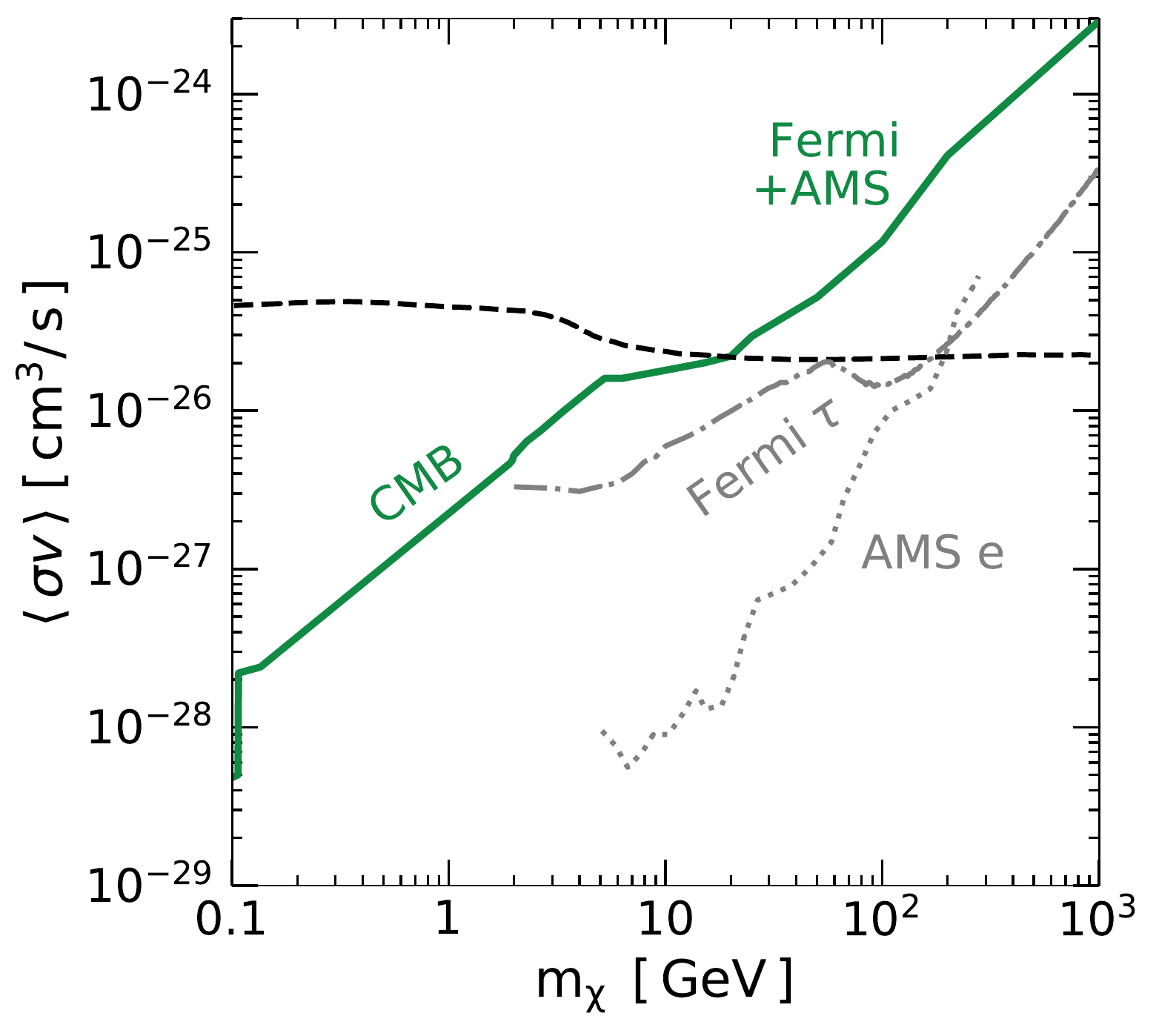}
\caption{Combined limit on the total annihilation cross section for WIMP dark matter, in the conservative case (solid). The bound on the total cross section at a given mass
is determined by the weakest combination of branching fractions. Also shown is the thermal relic line (dashed), and comparison with the standard $100\%$ cases for \textit{Fermi} $\tau$ (dot-dashed) and AMS electrons from Ref.~\cite{Bergstrom:2013jra} (dotted).}  
\label{fig:Comblimit}
\end{figure}
\begin{figure}[t]
\centering
\includegraphics[width=0.97\columnwidth]{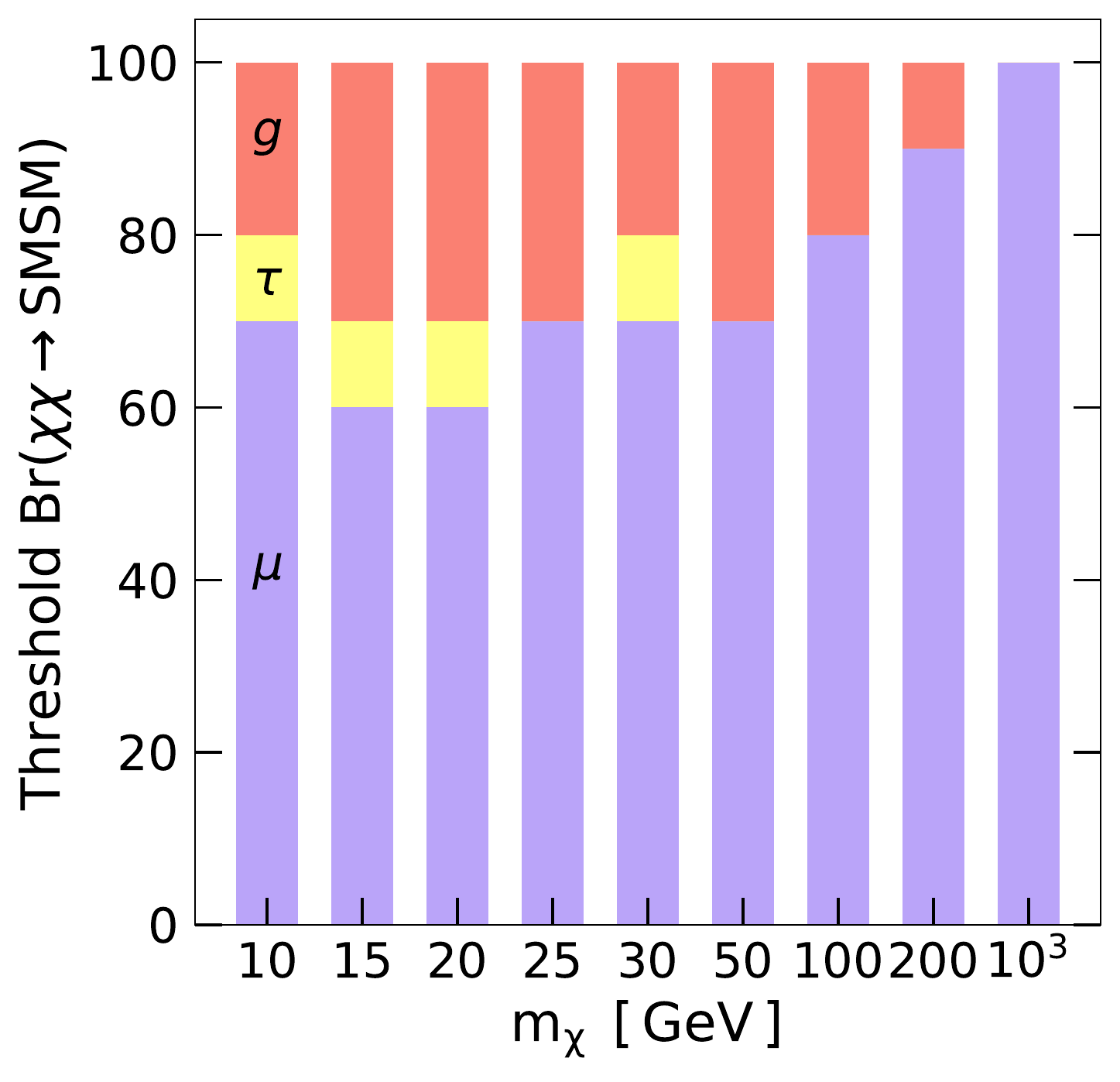}
\caption{Threshold branching fractions (least constraining combinations) that set the exclusion limit. We sample briefly at higher masses. Muons are the least constrained final state.}  
\label{fig:equalizer}
\end{figure}

Figure~\ref{fig:Comblimit} shows our calculation of the largely model-independent upper limit on the WIMP total annihilation cross section. We find the model-independent lower limit on the DM mass in the generic scenario is 20 GeV --- this is where the total cross section limit crosses the thermal relic line. Following the discussion in Sec.~\ref{sec:belief}, note that enforcing a clearance factor of 2 leads to a lower limit on the DM mass of about 6 GeV, and a clearance factor of 10 gives a lower limit of about 2 GeV.
Also shown for comparison are the standard 100$\%$ scenarios commonly considered in the literature, $\tau$ final states probed by \textit{Fermi}, and $e$ final states probed by AMS. It is clear that the general approach of comparing favorable single-channel limits with the thermal relic cross section badly overstates the degree to which thermal WIMPs have been probed.

At different DM masses, different experiments dominate the total limit. These regions are as follows:\\

\bm{$m_\chi \lesssim 135$}~\textbf{MeV:} In this region, the CMB is most constraining. The only available final states are muons, electrons, and photons. Therefore, we use the analytic spectra to obtain a limit, taking the least constraining of the three final states to find the combined limit. This region is excluded for generic WIMPs. The drop in the limit at $\sim105$~MeV, the muon mass, is the kinematic cutoff from a limit on muon final states, to electron and photon final states.\\

\bm{$0.135 \lesssim m_\chi \lesssim 5.1$}~\textbf{GeV:} In this region, the CMB is most constraining. For these masses, hadronic resonances introduce extra uncertainty in spectra generated with {\sc Pythia}. However, taking the conservative limit from Eq.~(\ref{eq:cmbhadronic}), the bound still remains below the relic line. The limit shown in this region is generated using {\sc Pythia}. This region is excluded for generic WIMPs.\\

\bm{$5.1 \lesssim m_\chi \lesssim 7$}~\textbf{GeV:} In this region, the CMB is most constraining. There are no hadronic resonances, and this limit simply comes from taking the least constrained final state from the CMB, taus, shown in Fig.~\ref{fig:CMBlimit}. As shown in Fig.~\ref{fig:Comblimit}, this region is below the thermal relic line, and so is excluded for generic WIMPs.\\

\bm{$7 \lesssim m_\chi \lesssim 1000$}~\textbf{GeV:} In this region, the combination limit from \textit{Fermi} and AMS is most constraining. This is shown as ``Fermi$+$AMS'' in Fig.~\ref{fig:Comblimit} (note the H.E.S.S. gamma-ray line search in this region prohibits large branching into gamma-ray line photons). This region is where the total cross section limit crosses the thermal relic line, giving a model-independent lower limit on the DM mass of $m_\chi\gtrsim$~20 GeV.\\

Figure~\ref{fig:equalizer} shows the threshold branching fractions for fixed masses, with DM masses binned with width 1 GeV. For the lower DM mass limit of 20 GeV, these are 60$\%$ to muons, $30\%$ to gluons, $10\%$ to taus, and 0$\%$ to the remaining final states. Note that the exclusion branching fraction combination is not necessarily unique. For many masses there are permutations of more than one final state that give a comparable limit (i.e., swap the gluon region with $b$-quarks, or some linear combination of $b$-quarks and gluons). The muon contribution however is generally present in all combinations, it is the smaller remaining combinations that vary. Muons are the least constrained among all the visible annihilation products.

\subsection{Quantifying ``Conservative''}
\label{sec:uncertainties}

Our limits are conservative; that is, we have consistently made choices that lead to a weaker final limit, within the parameter uncertainties. This is required to claim a meaningful robust lower limit on the DM mass. In this subsection, we detail steps taken to ensure a conservative limit for each experiment, and discuss variations from astrophysical uncertainties and modeling. 

\subsubsection{AMS}

\begin{figure}[t]
\centering
\includegraphics[width=\columnwidth]{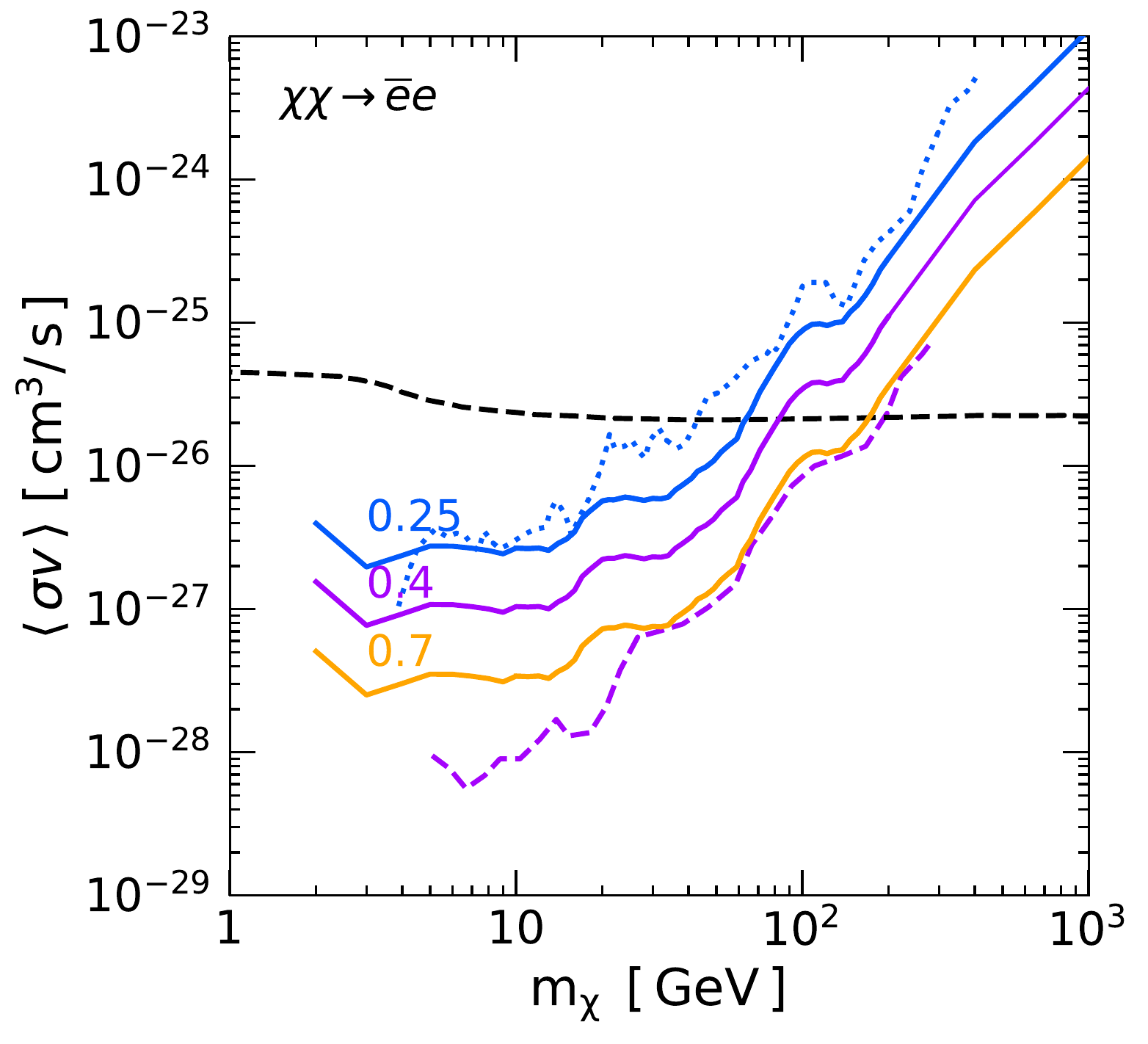}
\caption{Our conservative AMS limit illustrated for electron final states (solid), with a local DM density of 0.25 GeV/cm$^3$ (blue). The same conservative propagation parameter values, but with local DM densities of 0.4 GeV/cm$^3$ (purple) and 0.7 GeV/cm$^3$ (orange) are shown for comparison. As this is just a rescaling of results, the same range is found for all other final states. We also compare with the limits found in Ref.~\cite{Bergstrom:2013jra} (dashed, local DM density 0.4 GeV/cm$^3$) and Ref.~\cite{Cavasonza:2016qem} (dotted, local DM density 0.25 GeV/cm$^3$).}  
\label{fig:AMSuncertainty}
\end{figure}
\begin{figure}[t]
\centering
\includegraphics[width=\columnwidth]{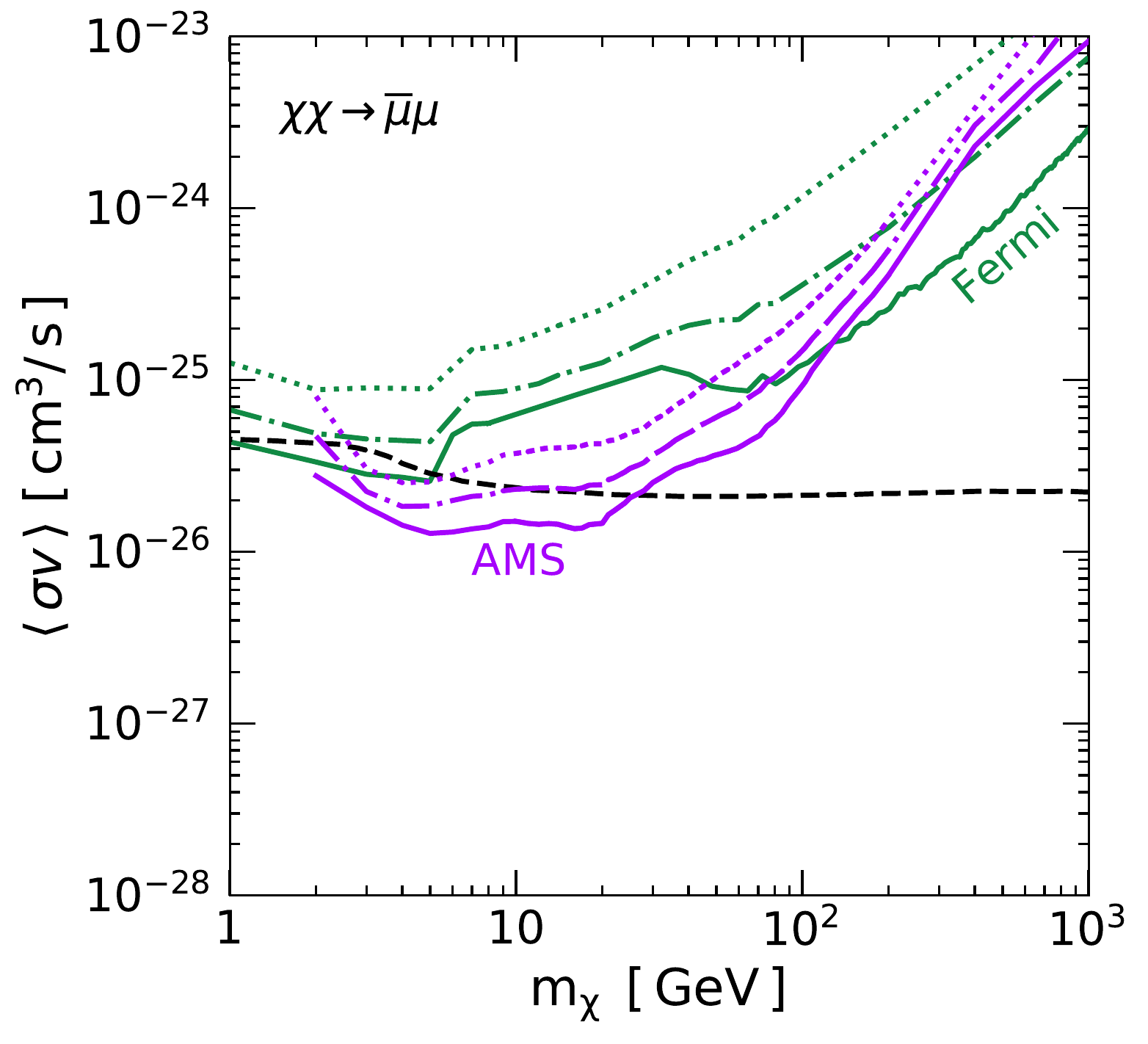}
\caption{AMS and \textit{Fermi} muon limits for  95$\%$ C.L. (solid), as well as 99.7 $\%$ C.L. (dot-dashed) and (1-10$^{-7})\times100\%$ C.L. (dotted) values. Thermal relic line is also shown (dashed).}  
\label{fig:statistics}
\end{figure}

Figure~\ref{fig:AMSuncertainty} compares our AMS limit for the $\chi\overline{\chi}\rightarrow\overline{e}e$ annihilation channel with the limit obtained in Ref.~\cite{Bergstrom:2013jra}. This shows our limit is more conservative -- i.e. weaker -- compared to existing bounds in the literature. The difference is because we have chosen parameter values (within the allowed ranges) that lead to weaker constraints, to ensure that any resulting exclusion will be robust. To reiterate and summarize our earlier discussion, in obtaining our limit we have made the following choices:

\begin{itemize}
 \item We do not model the backgrounds, and instead parameterize the total AMS measurements with a polynomial function. This does not require any assumptions about the interplay of background and signal propagation and their origins; we can take only the signal propagation as conservative. We check that this method produces a limit equally as weakly constraining as other works that do model both backgrounds and signals in a conservative manner~\cite{Bergstrom:2013jra}.
 \item We employ high energy losses by taking a conservative choice for magnetic fields, of $B_\odot=8.9 \, \mu$G at the Sun.
 \item We take the largest value of the solar modulation potential, $\Phi=0.6$ GV, measured for AMS during its data-taking period~\cite{Cholis:2015gna}. In Fig.~\ref{fig:AMSuncertainty}, the difference in our limit at low masses and other references is sourced by this choice.
 \item The local DM density is constrained to the range $\rho=$~[\,0.25,\,0.7\,] GeV/cm$^3$~\cite{Bovy:2012tw}. We take the lowest density of $\rho=0.25$ GeV/cm$^3$. This has the most dramatic impact on the limit, and is shown in Fig.~\ref{fig:AMSuncertainty}. Other choices such as propagation model, or choice of DM halo profile, have a sub-dominant effect on our result. 
\end{itemize}

These choices lead to a robust and conservative constraint within the framework of the standard diffusive propagation scenario for cosmic rays. If the propagation of electrons and positrons is substantially different than expected (e.g., \cite{Blum:2013zsa, Abeysekara:2017old}), this could impact our constraints; however, any such modification would need to obey stringent constraints from the wide range of cosmic-ray measurements at Earth (e.g., \cite{Profumo:2018fmz, Hooper:2017tkg}). Furthermore, our modeling of the unknown background with a smooth function should accommodate many modifications to the secondary flux of cosmic-ray positrons from astrophysical sources.

AMS also reports strong limits on DM annihilating to $b$-quarks from their antiproton dataset. These limits are stronger than \textit{Fermi} at low DM masses ($\lesssim$~50 GeV)~\cite{Reinert:2017aga}. We do not include the antiproton data in our combined limit. This is because the $b$-quark channel is not one of the key threshold channels; the weakest channels from each experiment are what sets the combined limit. Fig.~\ref{fig:equalizer} shows that the threshold does not contain $b$-quarks, and so introducing an additional analysis with stronger $b$-quark limits would not affect the limit on the total annihilation cross section.

Lastly, note that where AMS begins to lose sensitivity at low DM masses due to solar modulation, \textit{Voyager 1} begins to provide stronger limits~\cite{Krimigis1977, Stone150, Boudaud:2016mos}, as \textit{Voyager 1}
crossed the heliopause during data taking. We do not include \textit{Voyager 1} limits in our analysis, as the CMB is more constraining in this region.

As the AMS dataset has the largest uncertainties of all those considered in this paper, the choices discussed above have the greatest impact on the total annihilation cross section limit. Uncertainties from \textit{Fermi} and the CMB are comparably negligible. Regardless, we now discuss their relevant sources of uncertainty.

\subsubsection{Fermi}

The largest uncertainty from \textit{Fermi} is from the values of the $J$-factors. Taking a larger $J$-factor uncertainty of 0.8 dex does not change the \textit{Fermi} limit by more than about 10$\%$~\cite{Ackermann:2015zua}. The choice of DM halo profile leads to a negligible change, as the innermost region of the halo, where the density is most uncertain, does not dominate the limits.

Note that one of the dwarfs in the nominal sample, Tucana III, shows evidence of tidal stripping. This is a likely explanation of the reported excess of gamma rays over background, rather than DM. Excluding these systems strengthens the limit, but not substantially. Using the 2015 \textit{Fermi} analysis~\cite{Ackermann:2015zua} instead, which did not include any systems with excesses and only kinematically confirmed galaxies, increases the lower limit on the WIMP mass by only $\sim$few GeV. As this is clearly not a large effect, and to be most conservative, we choose to use the most recent full dataset which gives a weaker limit due to these excesses. (Note this low significance excess can be meaningfully combined with the AMS antiproton excess~\cite{Cuoco:2016eej,Cuoco:2017rxb,Eiteneuer:2017hoh,Jia:2017kjw,Arcadi:2017vis}.)

The case of DM annihilation in the Milky Way halo was considered recently in Ref.~\cite{Chang:2018bpt}, where stronger bounds were found for DM annihilating to $b$-quarks, compared to dSphs. However, similar to the AMS antiproton bounds, as Fig.~\ref{fig:equalizer} shows that the threshold branching combination does not contain $b$-quarks, introducing an additional analysis with stronger $b$-quark limits would not affect the limit on the total annihilation cross section.

\subsubsection{Planck}

The CMB bounds are the most robust and come with little theoretical uncertainty, especially as they do not depend on late time annihilation rates. In generating our CMB limits, the largest potential source of uncertainty comes from the energy spectra generated with {\sc Pythia}. However, in the less certain hadronic resonance regime we present arguments based on ionizing energy injection fractions in Fig.~\ref{fig:CMBenergyfrac}, and using Eq.~(\ref{eq:cmbhadronic}) confirm the limit is still below the thermal-relic line in this regime.

\subsubsection{Statistics}

Figure~\ref{fig:statistics} shows the variation of the AMS and \textit{Fermi} muon limits with higher statistical significance. We show the C.L. we take for our combined limit, 95$\%$ C.L. ($2\sigma$), as well as 99.7 $\%$ C.L. ($3\sigma$) and (1-10$^{-7})\times100\%$ C.L. ($5\sigma$) values.

\subsection{Implications for WIMP models}

We have studied a generic WIMP: an $s$-wave thermal relic with $2\rightarrow2$ annihilation to visible final states, with a standard thermal history and radiation dominated early universe. While this is a largely model-independent approach, the results have important implications for DM models. 

\subsubsection{Model Types}

Our approach covers models that have suppressed collider or direct detection signals. Important examples include scattering rates which are suppressed by powers of velocity or momentum, or cancellation between scattering diagrams~\cite{Ko:2016ybp, Bauer:2016gys, Duerr:2016tmh, Bell:2017rgi}. For Wino or Higgsino DM candidates, scattering occurs through suppressed loops~\cite{Hill:2013hoa}, but its annihilation is not suppressed.

The threshold branching fractions in Fig.~\ref{fig:equalizer} show that muons are the least constrained final state among all the visible annihilation products, and Fig.~\ref{fig:Comblimit} shows the combined limit is closest to following the AMS muon limit line for masses above 10 GeV, and the \textit{Fermi} muon limit line for masses above about 100 GeV. Large couplings to muons are possible within leptophilic DM models~\cite{Fox:2008kb, Kopp:2009et, Bai:2014osa, Bell:2014tta, Freitas:2014jla, delAguila:2014soa, DEramo:2017zqw}. Interestingly, even if DM does not couple to hadrons at tree-level, such interactions are induced at loop level, leading to hadronic contamination of the energy spectra~\cite{DEramo:2017zqw}. We do not include such effects in our spectra, but as these considerations are fairly generic, they could be used to make general statements about the WIMP.

\subsubsection{$2\rightarrow n$ Processes}

Compared to $2\rightarrow 2$ processes, $2\rightarrow n$ processes have energy spectra that are softened~\cite{Elor:2015bho,Elor:2015tva}. Therefore, such complications lead to even weaker lower limits on the DM mass compared the $2\rightarrow2$ case. This further supports the arguments in this work.

Our framework can be extended to hidden sector models with small DM-SM couplings, leading to an on-shell mediator $2\rightarrow 4n$ scenario, $\chi{\chi}\rightarrow(2n\times Y)\rightarrow(4n\,\times\,$SM), where $n$ is the number of cascade decays and $Y$ is a dark sector mediator. However, such a limit is only generically made assuming either the same mass mediator at each dark cascade step, or that the masses are much lower than the mass of their progenitor particle; otherwise the portion of DM energy split into each mediator's final states will be unequal~\cite{Bell:2016fqf,Bell:2016uhg}, introducing extra model dependence to the calculation.

Note that $2\rightarrow3$ bremsstrahlung processes can be the dominant DM annihilation mode in the scenario the $2\rightarrow2$ annihilation mode is suppressed~\cite{Bergstrom:1989jr,Flores:1989ru,Baltz:2002we,Bringmann:2007nk,Bergstrom:2008gr,Barger:2009xe,Bell:2010ei,Bell:2011if,Bell:2011eu,Ciafaloni:2011sa,Garny:2011cj,Bell:2012dk,DeSimone:2013gj,Ciafaloni:2013hya,Bell:2017irk,Bringmann:2017sko}. Bremsstrahlung can lift helicity suppression for direct annihilation for Majorana DM to neutrinos, but the annihilation rate is generally still not sufficiently large to produce a thermal relic cross section.

\begin{figure}[t]
\centering
\includegraphics[width=\columnwidth]{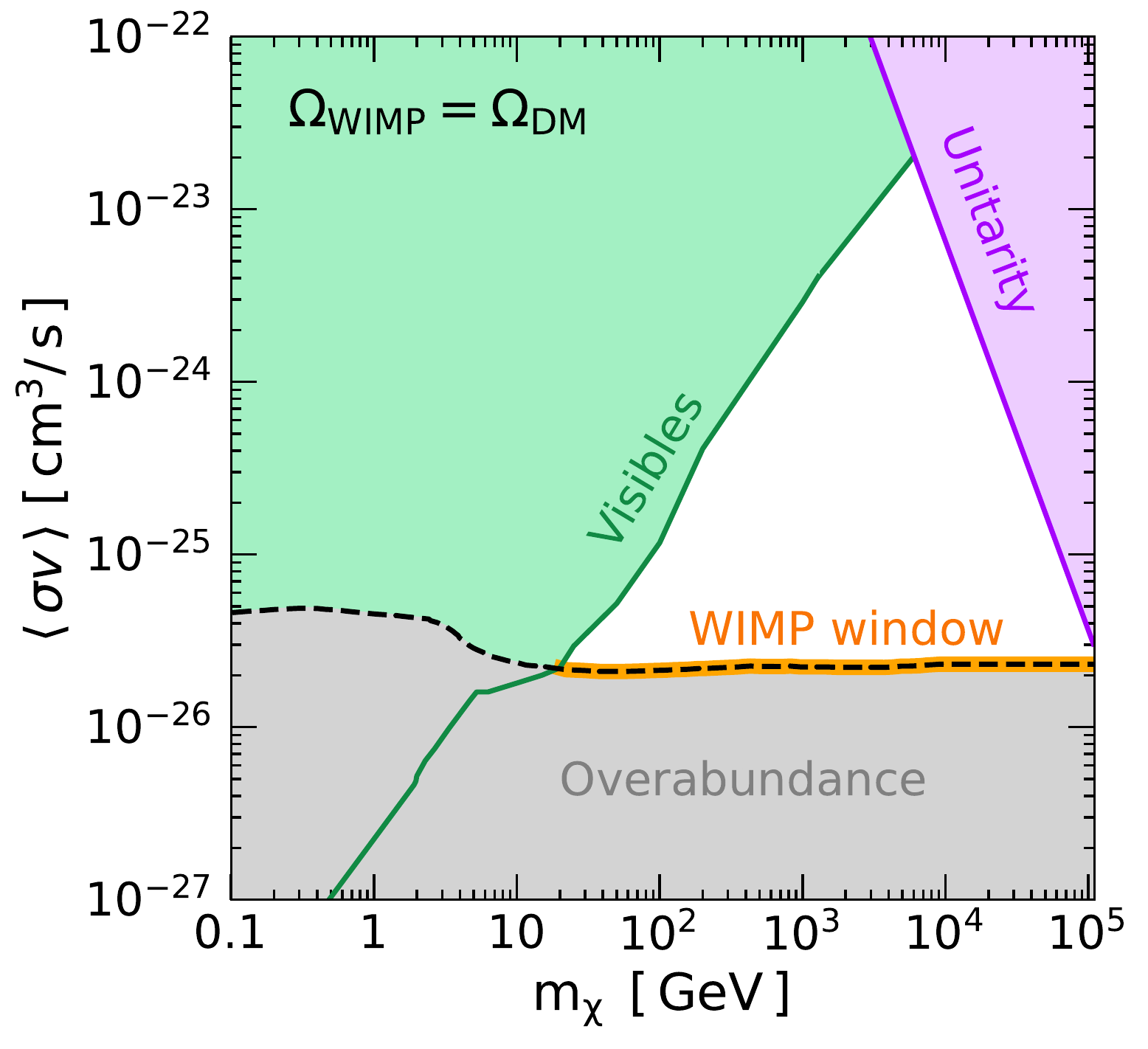}
\caption{Bounds on the generic thermal WIMP window (\mbox{$s$-wave} $2\rightarrow2$ annihilation,  standard cosmological history), assuming WIMP DM is 100$\%$ of the DM. Shown is the conservative bound calculated in this work from data (Visibles), and the unitarity bound~\cite{Griest:1989wd}. The remaining WIMP window is the orange line, and the white space is unprobed. Thermal relic cross section is the dashed line~\cite{Steigman:2012nb}.}  
\label{fig:window}
\end{figure}
\begin{figure}[t]
\centering
\includegraphics[width=\columnwidth]{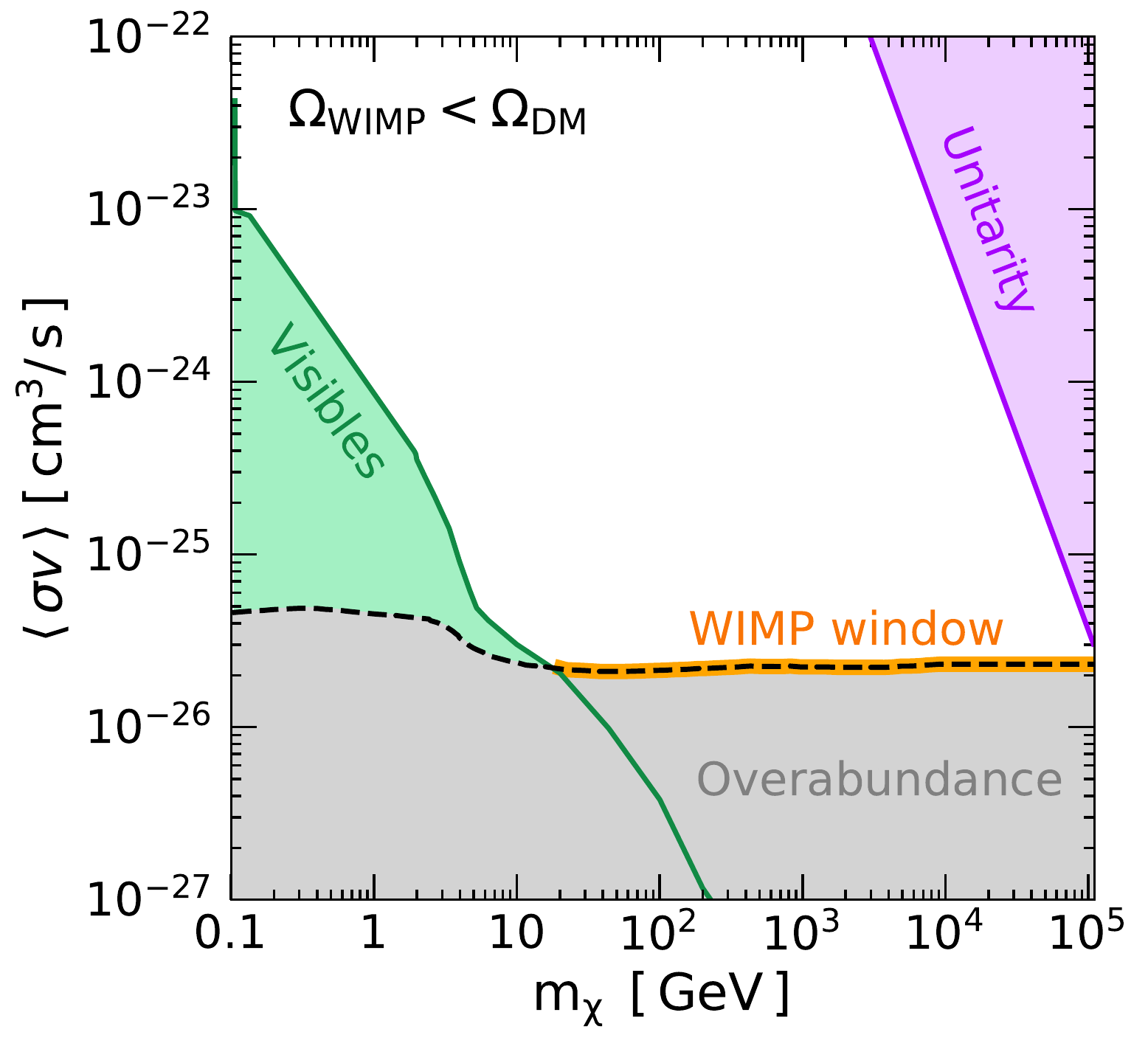}
\caption{Bounds on the generic thermal WIMP window (\mbox{$s$-wave} $2\rightarrow2$ annihilation,  standard cosmological history), assuming sub-dominant WIMP content. Shown is the conservative bound calculated in this work from data (Visibles), and the unitarity bound~\cite{Griest:1989wd}. Thermal relic cross section is the dashed line~\cite{Steigman:2012nb}.}  
\label{fig:windowsub}
\end{figure}

\subsubsection{Invisibles and Sub-Dominant Density}
\label{sec:subdom}
When the limit on the total cross section is below the thermal-relic prediction, the WIMP is nominally excluded.  There are two other possible interpretations. 

First, the fraction below the limit can be interpreted as the fraction required to proceed to invisible final states. 

Second, the strength of the limit below the relic line can also be used to set a bound on sub-dominant WIMP content. For standard indirect detection analyses for WIMP DM, the annihilation cross section and the density are often considered as independent, and are related to the astrophysical flux $F$ as
\begin{equation}
 F=\frac{\langle\sigma v\rangle}{8\pi m_\chi^2}\int\rho_\chi^2 d\ell,
\end{equation}
where $\rho_\chi$ is the DM density, and $\ell$ is the line of sight. The upper limit is obtained from upper limits on $F$, i.e.,
\begin{equation}
 \langle\sigma v\rangle < \langle\sigma v_{\rm limit}\rangle \equiv F\, \frac{8\pi m_\chi^2}{\int\rho_\chi^2 d\ell}.
\end{equation}
For sub-dominant WIMP DM, if the WIMP density is completely determined by the annihilation cross section, they are no longer independent, as
\begin{equation}
 \rho_{\rm WIMP}\langle\sigma v_{\rm WIMP}\rangle = \rho_\chi\langle\sigma v_\chi\rangle,
\end{equation}
where $\langle\sigma v_\chi\rangle\sim3\times10^{-26}$\,cm$^3/$s is the thermal relic cross section. The annihilation flux from the sub-dominant WIMP is then
\begin{align}
 F&=\frac{\langle\sigma v_{\rm WIMP}\rangle}{8\pi m_\chi^2}\int\rho_{\rm WIMP}^2 d\ell \nonumber\\
 &=\frac{\langle\sigma v_{\rm WIMP}\rangle}{8\pi m_\chi^2}\int\left(\frac{\sigma v_\chi\rho_\chi}{\langle\sigma v_{\rm WIMP}\rangle}\right)^2 d\ell \\
 &=\frac{\langle\sigma v_\chi\rangle^2}{\langle\sigma v_{\rm WIMP}\rangle}\frac{1}{8\pi m_\chi^2}\int\rho_\chi^2 d\ell. \nonumber
\end{align}
Therefore, an upper limit on the flux implies
\begin{equation}
 \frac{\langle\sigma v_\chi\rangle^2}{\langle\sigma v_{\rm WIMP}\rangle}<\langle\sigma v_{\rm limit}\rangle,
\end{equation}
which provides a lower limit on the sub-dominant WIMP cross section,
\begin{equation}
 \langle\sigma v_{\rm WIMP}\rangle > \frac{\langle\sigma v_\chi\rangle^2}{\langle\sigma v_{\rm limit}\rangle}.
 \label{eq:subdom}
\end{equation}

\subsection{The WIMP Window}

Figure~\ref{fig:window} illustrates the allowed mass range for a thermal WIMP, between our new general bound from Visibles (all SM states but neutrinos) and the bound from unitarity~\cite{Griest:1989wd}, assuming WIMPs are 100$\%$ of the DM. This window is $20\lesssim m_\chi \lesssim 1000$ GeV. Cross sections below the thermal-relic line are shaded as ``Overabundance'', as WIMPs with smaller cross sections produce more DM than observed, which is constrained with high accuracy~\cite{Patrignani:2016xqp}. The unitarity upper bound at larger masses can be escaped in non-standard scenarios, such as composite DM or in the presence of extra degrees of freedom, see i.e., Refs.~\cite{vonHarling:2014kha,Harigaya:2016nlg, Bramante:2017obj}. (Also note that in the presence of light mediators, contributions of higher partial waves to the cross section can weaken the unitarity constraint~\cite{Baldes:2017gzw}.) Note the bound shown in Fig.~\ref{fig:Comblimit} has been extended here --- for DM masses $m_\chi\gtrsim$~TeV, CMB bounds are strongest.

Figure~\ref{fig:windowsub} illustrates the allowed mass range for a sub-dominant WIMP, with a lower bound on the cross-section of WIMP DM with an arbitrary abundance. As the abundance is inversely proportional to $\langle\sigma v\rangle$, an upper limit on the astrophysical flux can be used to set a lower limit on $\langle\sigma v\rangle$ (see Eq.~(\ref{eq:subdom})). If WIMPs make up only part of the DM mass budget, its cross section is no longer restricted to be thermal. More generally, once the lower limit on the WIMP cross section exceeds the unitarity bound, fundamental WIMPs of this mass (DM or otherwise) will be totally ruled out~\cite{Blum:2014dca}.

\section{Towards Closing the WIMP Window}
\label{sec:outlook}

Future progress for decisively probing the WIMP paradigm requires improvement in indirect WIMP searches.

For \textit{Fermi}, the most important thing for progress is the discovery of new dwarf spheroidal galaxies, especially any that are closer to Earth. The Dark Energy Survey (DES) is well poised to achieve this goal~\cite{Drlica-Wagner:2015ufc, Diehl:2016ohr}, that could lead to an improvement in limits on the single channels by about an order of magnitude~\cite{Charles:2016pgz}. New, more powerful gamma-ray instruments with better angular resolution or greater sensitivity are needed in the $10+$ GeV range, such as \mbox{GAMMA-400}~\cite{2013arXiv1306.6175G,Cumani:2015ava} or HERD (High Energy cosmic-Radiation Detection)~\cite{Zhang:2014qga,Huang:2015fca}. Sub-GeV probes such as PANGU (PAir-productioN Gamma-ray Unit)~\cite{Wu:2014tya,Wu:2015wol}, AMEGO (All-sky Medium Energy Gamma-ray Observatory)~\cite{1748-0221-12-11-C11024}, or ComPair (Compton-Pair Production Space Telescope)~\cite{Moiseev:2015lva} will lead to better background subtraction for higher-energy searches.  Otherwise, the sensitivity reach for \textit{Fermi} will increase with the square root of exposure time. So far, \textit{Fermi} has collected nearly 10 years of data, but has reported constraints on dSphs using 6 years of data~\cite{Fermi-LAT:2016uux}. 

AMS, on the other hand, sets bounds using a shorter exposure time of $\sim2.5$ years~\cite{Aguilar:2014mma}, and so can expect a greater increase based on exposure time alone. A key issue facing AMS limits is CR background/propagation uncertainties, which substantially restrict the strength of our conservative limits. Working towards a better theory understanding would significantly aid progress for decisively excluding the WIMP (see also Ref.~\cite{Gaggero:2018zbd}). Experimentally, better measurement of the local DM density will have a substantial impact (see Fig.~\ref{fig:AMSuncertainty}), allowing a much stronger constraint on the lower DM mass. Analysis of new \textit{Gaia} data will be vital here~\cite{2016A&A...595A...2G,2018arXiv180409365G}.

Future probes of the CMB only expect an improvement of a factor of a few. This is due to a fundamental bound of cosmic variance: we only have one universe to measure.

There are good prospects for improvements at high mass. The Cherenkov Telescope Array (CTA) is often considered to be decisive for higher WIMP masses (above $\sim$100 GeV). While it will have an important role in testing WIMPs, it will not be able to close the current WIMP window. Progress is required below these masses first, to make decisive statements about the status of the WIMP. Other than CTA~\cite{Silverwood:2014yza}, current generation telescopes, H.E.S.S.~\cite{Abramowski:2011hc,Abramowski:2014tra,Profumo:2017obk,Abdallah:2018qtu}, VERITAS (Very Energetic Radiation Imaging Telescope Array System)~\cite{Archambault:2017wyh,Zitzer:2017xlo}, MAGIC (Major Atmospheric Gamma Imaging Cherenkov Telescopes)~\cite{Ahnen:2016qkx,Ahnen:2017pqx,Doro:2017dqn}, HAWC (High-Altitude Water Cherenkov Observatory)~\cite{Proper:2015xya,Abeysekara:2017jxs,Albert:2017vtb,Leane:2017vag,Ng:2017aur}, DAMPE (Dark Matter Particle Explorer mission)~\cite{TheDAMPE:2017dtc}, and in future LHAASO (Large High Altitude Air Shower Observatory)~\cite{DiSciascio:2015exu,DiSciascio:2016rgi}, will also aid eventually closing up to the unitarity limit at $\sim$100 TeV.

Improvements from neutrino searches may also be complementary~\cite{Beacom:2006tt,Yuksel:2007ac,Aartsen:2017ulx,ElAisati:2017ppn}, such as those from IceCube~\cite{Aartsen:2015xej}, ANTARES~\cite{Albert:2016emp}, and SuperK~\cite{1742-6596-718-4-042040}.

As muon-dominated final states are consistently the least constrained, anything that improves constraints on muon-rich models is well motivated by our study to close the thermal WIMP window. A recent study of DM annihilation in the Milky Way Halo~\cite{Chang:2018bpt} did not include inverse Compton scattering (ICS), which is sensible when being conservative, because the Galactic interstellar radiation field and magnetic field are not well understood. However, as a consequence, the constraints presented on muon-heavy models are not very strong. A similar approach with a careful accounting for ICS could set stronger constraints on muon-rich models. 

\section{Conclusions}
\label{sec:conclusion}

Recently, there has been some growing sentiment that thermal WIMPs are on death row. However, such statements are often based upon direct detection scattering rates, or collider missing-momentum searches. In both cases, there is no well-defined and predictive scale for WIMP-SM interactions, and only specific aspects of any such interactions are being probed. Furthermore, any interpretation of such limits requires model-dependent choices and additional assumptions that cannot be model-independently related to the WIMP thermal relic cross section. Such limits may exclude model-dependent possibilities, but reveal nothing about whether the remaining possibilities are viable or not. Probes of WIMP annihilation products are tied to the fundamental nature of the WIMP as an annihilation relic, and so allow for a decisive statement about the exclusion of the WIMP.

We have calculated the first largely model-independent upper limit on the total WIMP annihilation cross section from data, meaningfully combining bounds from all kinematically possible DM annihilation products. We find that thermal WIMPs with $s$-wave $2\rightarrow2$ annihilation to visible final states have a lower exclusion bound of \mbox{$m_\chi\gtrsim20$ GeV}. For the bound near 20 GeV, we have shown that the limit barely extends past the thermal relic scale, and pushing to higher statistical significance weakens this further. Enforcing a clearance factor of 2 leads to a lower limit on the DM mass of about 6 GeV, and a clearance factor of 10 gives a lower limit of about 2 GeV. For sub-GeV WIMP DM, the bound is severe, and extends to much lower masses.

The only way to decisively test the thermal WIMP scenario is to gain sensitivity to the total annihilation cross section down to the theoretical expectation. We have established an upper limit on the cross-section based on data, for the most generic WIMP scenario. This features an $s$-wave $2\rightarrow2$ annihilation cross section into visible final states, and a standard thermal history, with a radiation-dominated early universe. Other complications are possible, but they generally weaken limits. Considering a larger set of possibilities generally will just push the lower mass limit lower, which supports our point that GeV-scale WIMP DM is not even slightly dead.

We have discussed important improvements for moving towards closing the WIMP window. Discovery of new dwarf galaxies, with the aid of DES, would enhance limits from \textit{Fermi}. This could significantly improve sensitivity, leading with photon-rich final states. Refinement of AMS results through better understanding of CR propagation uncertainties will provide a better probe of leptonic final states. Together, using our framework, these experiments or similar have potential to model-independently exclude the generic WIMP up to the 100 GeV scale. Once this scale is reached, CTA will play a key role in excluding (or discovering) the WIMP. 
The remaining WIMP window is finite, and can ultimately be fully probed in a largely model-independent way.\\
 
\bigskip

\section*{Acknowledgments}

For helpful discussions, we are grateful to Iason Baldes, Nicole Bell, Fr\'ed\'eric Dreyer, Katherine Freese, Stefan Prestel, Leszek Roszkowski, Yotam Soreq, and Wei Xue. We also thank the anonymous referee for helpful comments. In addition to the software packages cited above, this research made use of {\sc IPython~\cite{Perez:2007emg}}, {\sc Matplotlib}~\cite{Hunter:2007}, and {\sc SciPy}~\cite{sicpy}. RKL and TRS are supported by the Office of High Energy Physics of the U.S. Department of Energy under Grant No. DE-SC00012567 and DE-SC0013999. JFB is supported by NSF Grant No. PHY-1714479. KCYN is supported by the Croucher Fellowship and Benoziyo Fellowship. TRS thanks the Kavli Institute for Theoretical Physics for its hospitality during the completion of this work; this research was supported in part by the National Science Foundation under Grant No. NSF PHY-1748958.

\bibliography{lowerDM.bib}

\end{document}